\documentclass[prd,tightenlines,superscriptaddress,nofootinbib,preprintnumbers,notitlepage]{revtex4-1}
\usepackage{amsmath,amssymb}
\usepackage{multirow}
\usepackage{graphicx}
\usepackage{tabularx}
\usepackage{slashed}
\usepackage{url}
\usepackage{hyperref}
\usepackage{rotating}
\usepackage{color,xcolor}
\usepackage{booktabs}
\usepackage[normalem]{ulem}

\newcommand{\arxiv}[1]{\href{http://arxiv.org/abs/#1}{arXiv:#1}}

\newcommand{\epem}{\ensuremath{e^+ e^-}}

\newcommand\one{\leavevmode\hbox{\small1\normalsize\kern-.33em1}}

\newcommand{\qqqquad}{\qquad \qquad \qquad}

\newcommand{\tev}{{\ensuremath\text{TeV}}}

\newcommand{\iab}{{\ensuremath\text{ab}^{-1}}}

\def\slashchar#1{\setbox0=\hbox{$#1$}           
   \dimen0=\wd0                                 
   \setbox1=\hbox{/} \dimen1=\wd1               
   \ifdim\dimen0>\dimen1                        
      \rlap{\hbox to \dimen0{\hfil/\hfil}}      
      #1                                        
   \else                                        
      \rlap{\hbox to \dimen1{\hfil$#1$\hfil}}   
      /                                         
   \fi}

\def\eg{\textsl{e.g.}}
\def\ie{\textsl{i.e.}}
\def\etal{\textsl{et al}}

\newcommand{\be}{\begin{eqnarray*}}
\newcommand{\ee}{\end{eqnarray*}}

\newcommand{\bee}{\begin{eqnarray}}
\newcommand{\eee}{\end{eqnarray}}
\newcommand{\beeq}{\begin{equation}}
\newcommand{\eeeq}{\end{equation}}

\begin{document}

\title{Higgs Factories: Higgs--Strahlung versus W--Fusion}

\author{R\'emi Lafaye}
\affiliation{LAPP, Universit\'e Savoie, IN2P3/CNRS, Annecy, France}

\author{Tilman Plehn}
\affiliation{Institut f\"ur Theoretische Physik, Universit\"at Heidelberg, Germany}

\author{Michael Rauch}
\affiliation{Institute for Theoretical Physics, Karlsruhe Institute of Technology (KIT), Germany}

\author{Dirk Zerwas}
\affiliation{LAL, IN2P3/CNRS, Orsay, France}

\begin{abstract}
  Higgs factories will be able to measure Higgs properties, and in
  particular Higgs couplings with high precision. In a follow-up to an
  earlier analysis we study the impact of the Higgs--strahlung and
  vector boson fusion processes on the precision of the couplings
  determination. Provided theoretical uncertainties can be controlled,
  future Higgs factories will be able to measure Higgs couplings at
  the sub-percent level. In spite of their different strengths,
  base-line circular and linear designs can expect a surprisingly
  similar performance.
\end{abstract}

\maketitle
\bigskip
\bigskip
\bigskip
\tableofcontents

\newpage
\section{Introduction}
\label{sec:intro}

The discovery of a light narrow Higgs boson~\cite{higgs} in
2012~\cite{discovery} has not only established the Higgs mechanism as
a key ingredient of the Standard Model, it has also provided a new,
powerful handle to test the structure of the Standard Model.  The
obvious question arises if this fundamental scalar sector really only
includes the single Higgs boson predicted by the Standard Model. ATLAS
and CMS performed a large number of tests of the observed Higgs
resonance, in which no significant deviations from the Standard Model
properties were observed, for example in the Higgs production and
decay
rates~\cite{ex_fits,legacy1,legacy2,legacy3,other_final_fits1,other_final_fits2}.
If we describe the Higgs couplings in terms of gauge-invariant
higher-dimensional operators we can relate the level of agreement with
the Standard Model predictions to the scale of new physics, for
example assuming weakly interacting models of new
physics~\cite{bsm_review}. In that case we find that the current
indirect limits from the Higgs sector are not significantly exceeding
the limits from direct LHC
searches~\cite{legacy2,legacy3,other_final_fits2}.  Translating for
example the measurements of anomalous gauge and Higgs couplings into
an underlying mass scale of an effective Lagrangian, weakly
interacting modifications of the Standard Model are only constrained
to be heavier than
$300~\dots~500$~GeV~\cite{legacy2,legacy3}. This energy scale is 
largely determined by the momentum flow through the relevant 
hard processes which was reached at Run~I. 
Significantly improving
this reach beyond the direct searches for example at the LHC is at the
heart of the case of an $\epem$ Higgs
factory~\cite{ilc_tdr,sfitter_ilc,michael_ilc}.\bigskip

The arguably most impressive achievement during the past LHC runs is
the extensive Higgs precision analysis program.  This is why one
cannot estimate the reach of a global Higgs analysis at future Higgs
factories without including the projected LHC results. This also means
that independent of the energy of a future Higgs factory there will be
a measurement of the top Yukawa coupling available for a global
analysis. However, a key problem of Higgs coupling measurements at the
LHC are the theoretical uncertainties. Higgs coupling measurements are
based on experimentally observed event rates, corresponding to cross
sections times branching ratios times efficiencies. These depend often
on the kinematic configuration of the events. Translating these
measurements into parameters of a Lagrangian requires QCD precision
calculations which only converge slowly for hadron collider
observables. Moreover, the more complex a signature, the harder it is
to even assign a reliable theoretical uncertainty to an
observation. Altogether, this means that Higgs coupling measurements
at the LHC will soon be limited by theoretical uncertainties which are
difficult to quantify and even more difficult to reduce. This means
that it will be very challenging to arrive at a conclusive statement
about possible new physics explanations, if a percent level deviation
from the SM-Higgs predictions should be observed.

\begin{figure}[b!]
\includegraphics[width=0.48\columnwidth]{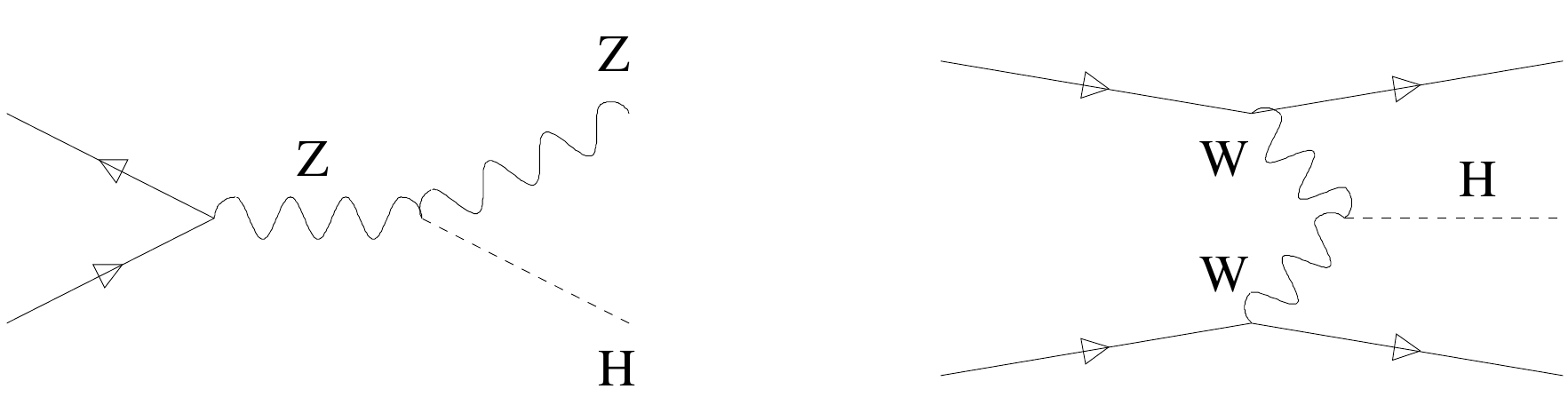} 
\caption{Feynman diagrams for Higgs--strahlung and $W$ fusion Higgs production channels at an
  $\epem$ Higgs factory.}
\label{fig:feyn}
\end{figure}

At an $\epem$ collider the two main Higgs production mechanisms are
Higgs--strahlung and $W$-fusion Higgs production~\cite{w_fusion}
\begin{align}
\epem \to ZH \qqqquad 
e^+ e^- \to \nu \bar{\nu} H \; ,
\end{align}
as illustrated in Fig.~\ref{fig:feyn}. They can be complemented by
associated production of top quark pairs and a Higgs boson and by
Higgs pair production in association with a $Z$-boson. Here the level
of uncertainties are very different~\cite{sfitter_ilc}: electroweak
corrections are typically a factor 10 to 100 smaller than QCD
corrections; while at a hadron collider large logarithms are for
example induced through the production of strongly boosted particles,
this is not an issue for a Higgs factory; analyses in an $\epem$
environment are less dependent on phase space configurations than at a
hadron collider and avoid the definition of fiducial phase space
regions; parton densities describing the incoming protons at the LHC
with sizeable uncertainties are replaced with very well understood
incoming electrons and positrons; initial state radiation of QCD jets
is replaced by better-controlled QED radiation and
Beam-strahlung. Finally, while at a hadron collider selected final
states of the Higgs boson decay are analyzed due to the overwhelming
background in some channels, e.g. for the Higgs decay to
gluons, an $\epem$ collider can analyze all standard final
states. This includes Higgs decays to charm quarks and gluons,
provided we take care in controlling the purity of the event
separation. All of this clearly points to an $\epem$ Higgs factory
to be built to test the nature of the Higgs boson.\bigskip

Two fundamentally different designs for Higgs factories exist to date:
the first design is a linear collider (ILC) planned with a
center-of-mass energy between 250~GeV and 500~GeV, in principle
upgradable up to 1~TeV. The ILC can run on the peak of the
Higgs--strahlung cross section as well as at larger
center--of-mass energy~\cite{ilc1,ilc2}. The latter option is particularly useful to
extract precise information from $W$-fusion Higgs production as well
as to measure Higgs production associated with top-quark pairs. One of
the aims of this study will be to quantify how much a Higgs couplings
analysis will gain from these higher-energy runs.
The linear design is also realized in the Compact Linear Collider
(CLIC) proposal~\cite{clic}, where the setup is particularly targeted to
operate at high energies up to 3~TeV, but will also include runs at
lower energies of 380~GeV and 1.5~TeV. At such large energies, a
consistent analysis should include a possible momentum-dependence of the
couplings in its setup, for example by using dimension-6 operators.
Current studies~\cite{clic,clichiggs} have considered only rate
measurements and no distributions, so we do not study the CLIC proposal
any further.

The second design is a circular $\epem$ collider that could produce
very large numbers of $ZH$ events close to
threshold~\cite{fcc-Higgs,fccinv}. Its strength is the large
luminosity due to the circular setup. However, a circular collider has
a limited center-of-mass energy below around 350~GeV.  In addition,
its luminosity decreases as function of energy, whereas it increases
for a linear collider.  We give the expected integrated luminosity
which we will use in our analysis in Tab.~\ref{tab:designs} based on
Refs~\cite{ilc2,ilc_private,fcc}. All values are subject to change as
staging, increased running time will impact the scenarios until
project approval (and after). The absolute values of the integrated
luminosities are not directly comparable, as the ILC is foreseen to be
run with polarized beams. The polarization has an impact on
the production cross section. While in the Higgs--strahlung process
left-chiral and right-chiral fermions couple to the $Z$--boson with
different coupling strengths, only left-chiral fermions induce
$W$-fusion Higgs production. Throughout the paper we will refer to the
circular collider as FCCee, implicitly including the CEPC design with
its slightly smaller integrated luminosities. The expected precisions given 
in~\cite{fcc-Higgs} do not scale as the square root of the luminosity between 
the FCCee and CEPC, we take the smaller of the two uncertainties. 

This discussion of the fundamental question whether optimizing a collider design
for  energy or for luminosity will lead to a better new physics reach is 
technically based 
on Ref.~\cite{sfitter_ilc}. We always assume that high-luminosity LHC
(HL-LHC) results for $3~\iab$ will be available for combination with
any of the Higgs factories.

\begin{table}[t]
\begin{tabular}{lrr}
\hline
collider         & $\sqrt{s}$ [GeV] & luminosity [$\iab$] \\\hline
HL-LHC           & 14000            & 3                   \\\hline
FCCee/CEPC-base  &   240            & 4                   \\
FCCee/CEPC-350   &   240/350        & 4/1                 \\\hline
ILC-stage        &   250            & 2                   \\
ILC-base         &   250/350/500    & 0.5/0.2/0.5         \\
ILC-lumi         &   250/350/500    & 2/0.2/4             \\\hline
\end{tabular} 
\label{tab:designs}
\caption{Design parameters for different Higgs factories taken from
  Refs.~\cite{ilc2,ilc_private,fcc}. Note that the luminosities are
  not directly comparable, as the ILC design includes polarized
  beams.}
\end{table}

\section{SFitter setup}
\label{sec:setup}

The setup of this updated analysis closely follows the original
analysis of Ref.~\cite{sfitter_ilc}.  Higgs couplings are defined as
prefactors of the respective Lagrangian terms coupling the Higgs field
to other Standard Model particles~\cite{sfitter_higgs,duehrssen},
\begin{align}
g_{xxH} \equiv g_x  = 
\left( 1 + \Delta_x \right) \;
g_x^\text{SM} 
\qquad \text{and} \qquad 
\frac{g_x}{g_y} = \left( 1 + \Delta_{x/y} \right)
\frac{g_x^\text{SM}}{g_y^\text{SM}} \;.
\label{eq:delta}
\end{align}
The loop-induced Higgs-photon (Higgs-gluon) coupling can be modified
two ways, through shifted, underlying tree-level Higgs couplings to
Standard Model particles and through new particles in the loop,
\begin{align}
g_{\gamma\gamma H} &\equiv g_{\gamma}  = 
\left( 1 + \Delta_\gamma^\text{SM} + \Delta_\gamma \right) \;
g_\gamma^\text{SM} \; 
= \left( 1 + \Delta_\gamma^\text{SM+NP} \right) g_\gamma^\text{SM} \;.
\label{eq:deltagamma}
\end{align}
Equivalent parameters
$\kappa_x \equiv 1+\Delta_x$ have been introduced in
Ref.~\cite{HiggsXS}, without disentangling modified tree-level
couplings and new states in the loop-induced couplings.  For
deviations of the Higgs width from its SM value we additionally
define
\begin{align}
\Delta_\Gamma &= \frac{\Gamma_\text{tot}(H) -
\Gamma_\text{tot}^\text{SM}(H)}{\Gamma_\text{tot}^\text{SM}(H)} \;.
\end{align}
The one Higgs coupling we do not comment on is the self-coupling.
Measuring the Higgs self-coupling is a challenge at any
collider~\cite{LHCHiggsself, fcc-Higgs, ilc2}. For a center-of-mass
energy above 450~GeV, \ie\ for the ILC setup only, a determination with
30\% accuracy at the end of the high-luminosity stage seems
feasible~\cite{ilc2}. While its measurement might serve as a test of the
structure of a SM-like Higgs potential, it is not obvious where it would
give us additional information about relevant new physics
models.\bigskip

A Lagrangian based on the shifted Higgs couplings of
Eq.\eqref{eq:delta} is perfectly fine after electroweak symmetry
breaking, once we include a renormalization prescription. The problem
is the electroweak renormalizability of the Higgs sector, because this
condition fixes each Higgs coupling exactly to its SM prediction.
Certainly at a Higgs factory electroweak corrections will be crucial
to extract a Higgs coupling measurement~\cite{passarino}.  Modified
Higgs couplings can be properly defined in different ways. One way is
based on a renormalizable ultraviolet completion of the free Higgs
couplings model. In such a case we can link the shift in the light
Higgs couplings to the well-defined parameters of an ultraviolet
completion, like for example an extended Higgs
sector~\cite{sfitter_uv}.

Second, we can systematically construct an effective dimension-6
Lagrangian of the Higgs sector and identify essentially all $\Delta_x$
with Wilson coefficients $c_x/\Lambda^2$ in the
non-linear~\cite{nonlinear,legacy3} representation of the Higgs and
Goldstone fields. While in the pure Higgs sector at the LHC the linear
and non-linear results can be related through re-mapping of the
operator basis~\cite{legacy3} this is no longer true for the combined
Higgs and gauge sectors.  At the LHC, additional momentum-dependent
couplings play a major role in searching for deviations from Standard
Model kinematics, but at a Higgs factory we expect these additional
operators not to dominate our analysis. To be compatible with other
Higgs factory studies we stick to the modified Higgs couplings or
non-linear notation.

All linear collider measurements used in this study are taken from
Refs.~\cite{michael_ilc,AguilarSaavedra:2001rg,ilc_tdr}, with the
exception of the measurement of the $W$-fusion process with a $H \to
b\bar{b}$ decay at 250~GeV~\cite{Deschprivate}. The statistical
uncertainties are scaled to the corresponding integrated luminosity.
The expected error on the luminosity measurement of
0.3\%~\cite{BozovicJelisavcic:2010hy} is added to each measurement,
taken as fully correlated between them.  For the $\epem$ collider
production cross sections we assume a theoretical uncertainty of 0.5\% for
the $ZH$ and $W$-fusion production processes and 1\% for $t\bar{t}H$
production. Unlike for the LHC, at an $\epem$ Higgs factory the only
limiting theoretical uncertainty is on the partial width $\Gamma (H
\to b\bar{b})$, propagated into the different branching
ratios. Whenever we include this theoretical uncertainty we estimate
the error on the Higgs branching ratio to be around 2.2\%,
consisting of 0.65\% due to missing higher orders, 0.73\% from the
quark mass measurement(s), and 0.79\% from the value of
$\alpha_s$~\cite{HXSWGYR4,hdecay}, otherwise all these contributinos are
set to zero. This is consistent with our earlier
estimate of a combined 4\% uncertainty for the leading partial width
$\Gamma (H \to b\bar{b})$~\cite{sfitter_ilc}.\bigskip

One well-motivated assumption which is crucial for our analysis is
that a Higgs factory is expected to operate after the LHC has analyzed
a sizable Higgs data set.  Our conservative HL-LHC projections are
based on the previous detailed studies~\cite{sfitter_higgs,duehrssen}.
They include the main production and decay channel combinations, which
are expected to be measured with $3~\iab$ of integrated luminosity at
14~TeV center-of-mass energy, using results from a single
experiment. Since background systematics play an important role at
this high-luminosity result, the statistical gain from combining both
major experiments is expected to be much smaller than a naive scaling
with luminosity.  The two LHC channels, which will be most relevant
for our $\epem$ study, are inclusive Higgs production with decay into
a photon pair and top-quark-associated Higgs production with a Higgs
decay to photons. This comparably clean final state makes it a leading
channel to directly probe the top Yukawa coupling, despite the rather
low numbers of events originating from an inclusive cross section just
above a femtobarn.

For the LHC we scale all statistical errors to the increased
integrated luminosity.  The statistical component of experimental
uncertainties on background rates, which are determined from data,
will improve correspondingly.  The increase of statistics will also
improve the statistical component of the systematic errors. On the
other hand, experimental conditions (pile-up) will become
significantly more difficult for some of the crucial channels, like
$W$-fusion Higgs production and hadronic Higgs decays.  Therefore, the
same performance of particle identification and $b$-tagging as for
lower instantaneous luminosity is assumed, \ie\ the relative errors
for experimental systematics used in the previous studies are not
changed.  Theoretical errors on the cross sections and on the Higgs
branching ratios are taken from Ref.~\cite{HXSWGYR4} and included via
the profile likelihood \textsc{Rfit} scheme~\cite{rfit,sfitter},
implying that they are added linearly. Because Higgs analyses start to
depend more on exclusive jet observables~\cite{jetveto}, rendering the
application of fixed-order QCD corrections difficult, we refrain from
postulating an improved theoretical uncertainty.\bigskip

Unless we want to limit our LHC analysis with an overwhelming
universal error source, we have to make an assumption about the total
Higgs width. We generally assume
\begin{align}
\Gamma_\text{tot} = \sum_\text{obs} \; \Gamma_x(g_x) 
+ \text{2nd generation} \; .
\label{eq:width}
\end{align}
At $\epem$ colliders the total width can be inferred from a
combination of measurements. This is mainly due to the measurement of
the inclusive $ZH$ cross section based on a system recoiling against a
$Z \to \mu^+\mu^-$ decay.  While the simultaneous fit of all couplings
will reflect this property, we can illustrate this feature also
schematically. 

The simpler of two possibilities relies on two
measurements,
\begin{enumerate}
\item Higgs--strahlung inclusive: $\sigma_{ZH} \propto g_Z^2$\,,
\item Higgs--strahlung with a decay to $ZZ$: $\sigma_{ZH} \times \text{BR}_{ZZ} \propto g_Z^4/\Gamma_\text{tot}$\,.
\end{enumerate}
As the Higgs branching ratio into $ZZ$ is small, this requires a
significant amount of luminosity.  Another way to determine the Higgs
width is based on four
measurements~\cite{AguilarSaavedra:2001rg,ilc_tdr} of both production
processes shown in Fig.~\ref{fig:feyn},
\begin{enumerate}
\item Higgs--strahlung inclusive: $\sigma_{ZH}$\,,
\item Higgs--strahlung with a decay to $b\bar{b}$: $\sigma_{Zbb}$\,,
\item Higgs--strahlung with a decay to $WW$: $\sigma_{ZWW}$\,,
\item $W$-fusion with a decay to $b\bar{b}$: $\sigma_{\nu\nu bb}$\,,
\end{enumerate}
involving the four unknown parameters $\Delta_W$, $\Delta_Z$,
$\Delta_b$, and $\Gamma_\text{tot}$.  Schematically, the total width
can be extracted as~\cite{Deschprivate}
\begin{align}
\Gamma_\text{tot} \leftarrow 
\dfrac{\sigma_{\nu\nu bb}/\sigma_{Zbb}}
      {\sigma_{ZWW}/\sigma_{ZH}}
\times \sigma_{ZH} \; .
\end{align}
\bigskip

A final assumption at the LHC concerns hadronic Higgs decays.  Going
back to the width definition in Eq.\eqref{eq:width}, at the LHC the
Higgs decay to charm quarks with Standard-Model-like coupling
strengths is experimentally challenging. On the other hand, it
contributes to the total width at the per-cent level.  This is why we
usually link the second generation to the third generation via
\begin{align}
g_c = \frac{m_c}{m_t} \; g_t^\text{SM} (1 + \Delta_t) \; .
\label{eq:charm}
\end{align}
The leptonic muon Yukawa might be
observable at the LHC in weak boson fusion or inclusive searches,
depending on the available luminosity~\cite{muons}.  At an $\epem$ Higgs
factory such a link is not needed.

\section{Energy versus luminosity}
\label{sec:analysis}

\begin{figure}[b]
\includegraphics[width=0.8\columnwidth]{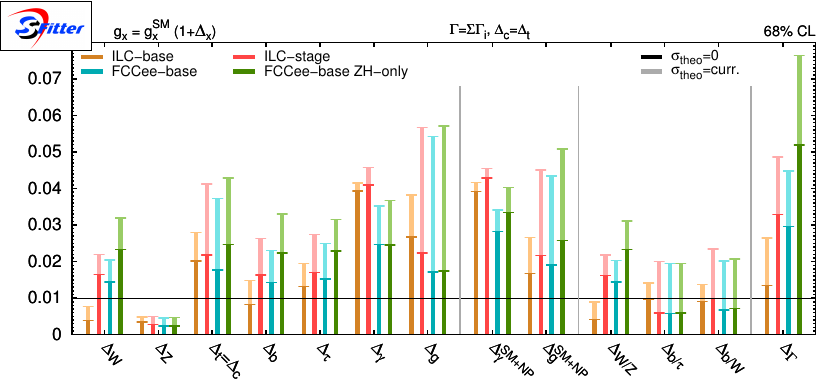}
\caption{Precision of the Higgs couplings extracted in the linear and
  circular baseline scenarios using the current theoretical errors and
  assuming negligible theory errors. We also show results assuming a
  staged low-energy operation of the ILC and the impact of the
  $W$-fusion process by restricting the FCCee measurements to $ZH$
  production. We assume that the total Higgs width is constructed from
  all observed partial widths.}
\label{fig:base}
\end{figure}

We start our comparison of the impact of the different experimental
channels on the Higgs couplings analysis following the base scenarios
defined in Tab.~\ref{tab:designs}.  In Fig.~\ref{fig:base} we show the
projected coupling measurements, assuming that the total width can be
computed as the sum of all observed partial widths plus contributions
from second-generation fermions, as defined in Eq.\eqref{eq:width}. In
particular, we assume that the charm Yukawa scales with the top Yukawa
following Eq.\eqref{eq:charm}.  In our comparison of FCCee and ILC
projections we always implicitly assume a combination with the
expected HL-LHC results.  Unless we make a significant theory
effort~\cite{theory_future}, the linear and circular designs will both
be limited by the theoretical uncertainty in extracting Higgs couplings
from observed event rates.  Assuming that this will eventually change,
we always show results for both the current accuracy and when ignoring
theoretical uncertainties for the $\epem$ colliders in lighter and darker
bands.\bigskip
 
The general pattern we see in Fig.~\ref{fig:base} is that for all
couplings with the exception of $\Delta_W$ the FCCee base design, the
ILC base design, and the ILC staging design give comparable
results. As we can see in Tab.~\ref{tab:designs}, the FCCee based
design and the initial staging ILC design are similar. This can be understood
from the impact of the beam polarization which increases the effective cross
section. However, for the full ILC design this outcome is not at all
trivial, given that the FCCee foresees eight times the integrated luminosity at
the lowest energy, neglecting beam polarization effects. The lower
integrated luminosity (total $1.2~\iab$) is compensated by the higher
center--of--mass energy. The only major difference occurs for $\Delta_W$. Here,
the statistical uncertainty on $ZH$ production with a subsequent decay
$H \to WW$ in the FCCee setup is around 0.9\%, but the larger error
on the indirectly obtained Higgs width limits the achievable precision
on $\Delta_W$. We confirm this by only including the $ZH$ measurements
in the FCCee base setup. In this case mostly the worse Higgs width
estimate increases the error on $\Delta_W$ and other couplings,
illustrating how even the FCCee at 240~GeV benefits from $W$-fusion.

The ILC will probe the $W$-fusion process combined with all main decay
modes, allowing for a higher overall precision. This
advantage will be reduced once we include electroweak precision
measurements testing the custodial symmetry enforcing
$\Delta_W=\Delta_Z$. Assuming custodial symmetry also improves the
precision on the Yukawa couplings, in particular when we ignore theory
uncertainties. For example, the uncertainty on $\Delta_b$ reduces to
2.8\% with current theoretical errors and 1.0\% without. On the other
hand, the accuracy on new-physics contributions to the loop-induced
couplings and ratios of couplings is hardly affected by this additional
constraint.

Minor differences between the scenarios are the slightly better
expected performance of the FCCee for $\Delta_\gamma$, and the
sizeable impact of theoretical uncertainties on the FCCee measurement of
$\Delta_g$. These theoretical uncertainties are the main source of
uncertainty for the branching ratio into both gluons and charm
quarks. The direct measurement of Higgs decays to charm quarks also
determines the top Yukawa coupling more precisely than the HL-LHC
$t\bar{t}H$ measurements, if we indeed set $\Delta_c=\Delta_t$.  When
deriving $\Delta_g$, the theoretical uncertainties add up, leading to
rather large error bars. When these are set to zero, the total width
entering the branching ratios becomes the main source. As this induces
a positive correlation between the different decay modes, $\Delta_g$
can now be determined more precisely than $\Delta_c$ and
$\Delta_g^\text{SM+NP}$, which parametrizes effects on the full
effective $Hgg$ coupling, leading to this large improvement.\bigskip

\begin{figure}[b!]
\includegraphics[width=0.45\columnwidth]{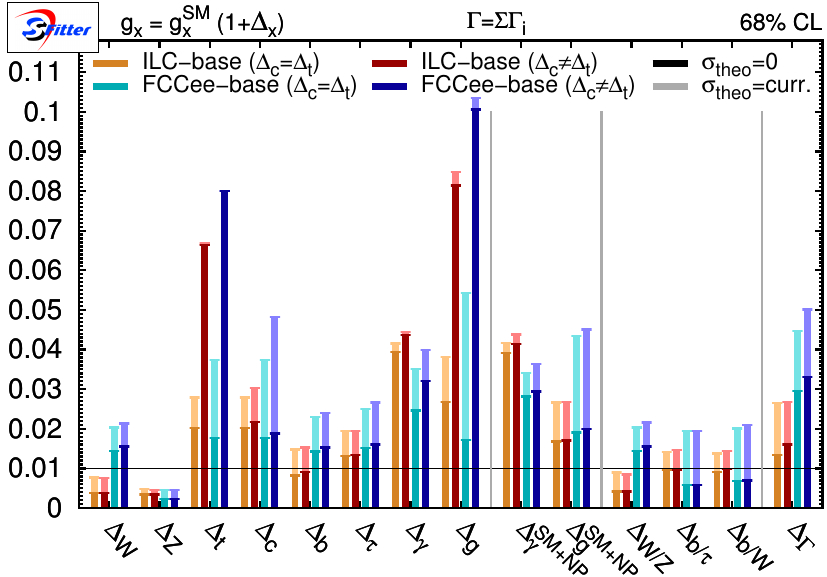} 
\hspace*{0.05\columnwidth}
\includegraphics[width=0.45\columnwidth]{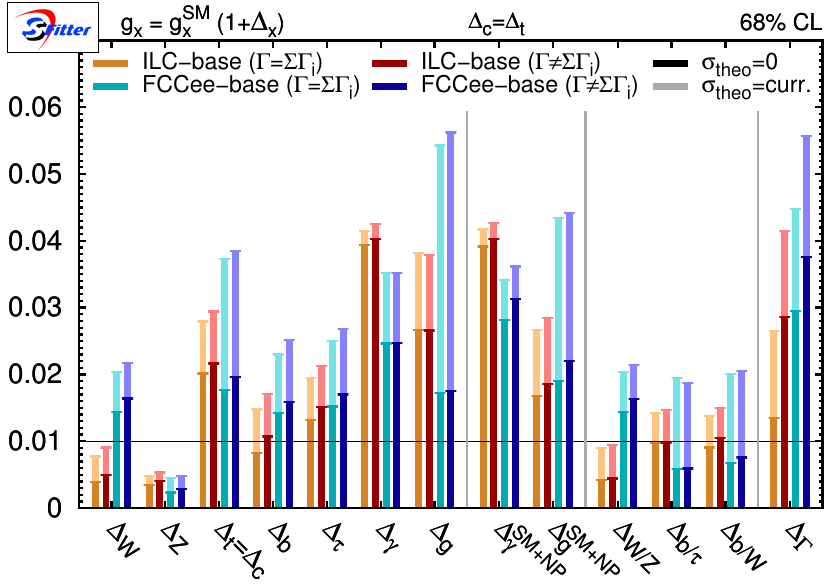}
\caption{Higgs coupling measurement for relaxed model assumptions in
  the linear and circular baseline scenarios. In the left panel we
  allow for $\Delta_c \ne \Delta_t$. In the right panel we include a
  direct Higgs width measurement in the presence of unobservable Higgs
  decay modes.}
\label{fig:relax}
\end{figure}

In a second step, we test the impact of the two model assumptions
we made in our initial estimate, because they are essentially
motivated by the type of measurements available at the LHC.  In the
left panel of Fig.~\ref{fig:relax} we show the effect for the two
base designs if we separate the charm and top Yukawas instead of using
Eq.\eqref{eq:charm}. The corresponding signature is Higgs decays $H
\to c\bar{c}$, which can be tagged very well in the clean $\epem$
environment. The fact that for both collider designs $\Delta_c$ is
essentially unchanged implies that most of the combined $\Delta_t$ and
$\Delta_c$ measurement comes from the charm Yukawa. In contrast, the
precision on $\Delta_t$ is reduced by almost a factor three. For the
FCCee setup, the determination of $\Delta_t$ now relies solely on the
HL-LHC results, while the 500~GeV run of the ILC just above the
$t\bar{t}H$ threshold yields a second measurement of similar accuracy.
Because the top loop gives a large contribution to the loop-induced
couplings means that a poorer measurement of $\Delta_t$ translates
into a poorer measurement of $\Delta_\gamma$ and $\Delta_g$. If we
combine all information on loop-induced decays into one coupling
deviation, $\Delta_{\gamma, g}^\text{SM+NP}$ remains unchanged in both
cases. At the same time, we also remove the link between the second
and third generation in the leptonic sector, allowing $\Delta_\mu \ne
\Delta_\tau$. However, the small Higgs branching ratio into muons in
the Standard Model prevents this channel from having an effect on our
global parameter analysis.

In the right panel of Fig.~\ref{fig:relax}, we relax the assumption on
the total width given in Eq.\eqref{eq:width}, allowing for additional
positive contributions. 
Technically, these can be invisible modes or decay modes which generate final
states of Standard Model particles different from the standard Higgs decay
channel analyses.
Because sizeable contributions to the total width also
mean sizeable event rates, for example invisible Higgs decays will be
strongly constrained by targeted searches.  If a deviation in the
total Higgs width cannot be mapped to a single signature, the direct
measurement will become difficult and our analysis will benefit from a
measurement of the total width.  For the FCCee the precision of this
actual Higgs width measurement is driven by the first scheme described
in Sec.~\ref{sec:setup}, combining the measurement of the inclusive
$ZH$ cross section with the $ZZ$ decay channel. The uncertainty on the
total width ranges from 5.6\% to 3.8\%, depending on the assumed
theoretical errors.  For the ILC the second scheme, including the
$W$-fusion production mode, dominates. It leads to a precision between
4.1\% and 2.9\%, respectively. Ignoring theoretical uncertainty the
FCCee setup matches the expected ILC base setup when we increase the
integrated luminosity by a factor 1.7.  Interestingly, we also find
that only using a direct measurement of the total width leads to very
little degradation in our heavily correlated global coupling
measurement, independent of slight differences in the actual Higgs
width determination.\bigskip

Going back to the results of Fig.~\ref{fig:base} we can see that even
ignoring theory uncertainties the difference between the energy-driven
ILC base scenario and the luminosity-driven FCCee base scenario is not
large.  Obviously, a staged ILC at 250~GeV is very similar in
performance to the FCCee base scenario, but in comparing the
luminosities the beam polarization has to be taken into account. The
most significant difference between the energy-driven and
luminosity-driven scenarios is the measurement of the Higgs-$W$
coupling, where the $W$-fusion process at higher energies leads to an
improvement of the measurement by up to a factor four, but deviations
between the Higgs-$W$ and Higgs-$Z$ couplings are also strongly
constrained by electroweak precision data. Model assumptions in
defining our hypothesis hardly play any role, so we can as well
minimize them. This in particularly relevant for any assumption on the
Higgs width, which is crucial at hadron colliders and simply not
needed at an $\epem$ Higgs factory.

\section{Converging upgrades}
\label{sec:upgrades}

\begin{figure}[t]
\includegraphics[width=0.8\columnwidth]{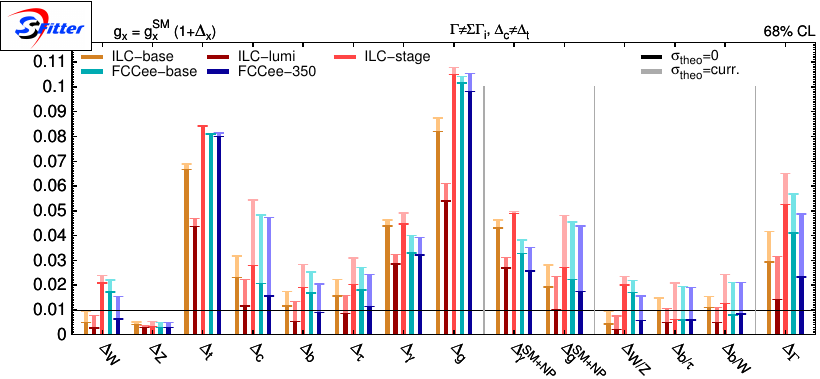} 
\caption{Precision of the Higgs couplings, allowing $\Delta_c \ne
  \Delta_t$ and unobserved decay channels. The Higgs width measurement
  does not include a model hypothesis. We show the FCCee and ILC
  baseline scenarios as well as their luminosity upgrades.}
\label{fig:upgrade} 
\end{figure}

While the FCCee and ILC base design are defined through their focus on
luminosity or energy in measuring Higgs couplings, we see in
Tab.~\ref{tab:designs} that the respective upgrade strategies will be
much more similar to each other. Compared to the base design, an
upgrade of the circular collider option targets the limitations in
energy, increasing the center--of--mass energy to 350~GeV. 
At the same time, the envisioned ILC upgrade will
focus on an increased integrated luminosity, at the 250~GeV low-energy
run and at the 500~GeV high-energy run.\bigskip

In Fig.~\ref{fig:upgrade} we show the expected precision for all Higgs
couplings when we relax both unnecessary conditions on the Higgs width
and on the second-generation Yukawa couplings, as studied in
Fig.~\ref{fig:relax}. If we focus on the experimental performance and
ignore the theory uncertainties, most Higgs couplings will be measured
below the percent level at the upgraded Higgs factories. The limiting
factor will most likely be the top Yukawa coupling, affecting the
higher-dimensional Higgs coupling to photons and gluons. For the
combination $\Delta_{\gamma, g}^\text{SM+NP}$, which is fixed directly
from the corresponding branching ratios, we also find projected
measurements of a few per-cent.

The differences between the upgraded ILC and the FCCee designs are
relatively minor: for $\Delta_W$ the ILC with its multitude of
$W$-fusion signatures will reduce the errors by a factor two to around
0.27\%. In contrast, for the luminosity-driven $\Delta_Z$ measurement
the FCCee upgrade will reach a precision of 0.28\%, limited by the
systematic error on the luminosity measurement. If this uncertainty
can be improved, statistics allows for a determination of this
coupling below the per-mill level. At the ILC the limit will be
slightly weaker.  The situation is reversed for the top Yukawa
coupling, where FCCee will not be able to improve the bounds obtained
from the HL-LHC run, while the accuracy with the upgraded ILC will 
beat the HL-LHC by more than a factor two. This translates directly
into improved bounds on dimension-5 contributions to the effective
$Hgg$ and $H\gamma\gamma$ couplings, parametrized by $\Delta_g$ and
$\Delta_\gamma$, respectively, and for the full new-physics effects,
$\Delta_{\gamma, g}^\text{SM+NP}$. Due to the higher statistics of
FCCee the difference is smaller for the parameters describing the
effective Higgs-photon coupling.

If we will not be able to improve on the theory uncertainties, the
errors on the coupling determination roughly double in size for most of
the tree-level Higgs couplings, such that only $\Delta_W$ for the ILC
upgrade and $\Delta_Z$ stay below the 1\% mark. The only exception is
the top Yukawa coupling, where the statistical component clearly
dominates and the theoretical uncertainties yield only a small
additional contribution. This then translates into the same situation
for the dimension-5 contributions to the effective $Hgg$ and
$H\gamma\gamma$ couplings, which also worsen only mildly.

Finally, we note that using ratios of couplings allows for
better probes for an upgraded FCCee with reduced theoretical errors, as
one can see for $\Delta_{b/\tau}$ and $\Delta_{b/W}$ in
Fig.~\ref{fig:upgrade}. The reason is that in this case the
precision on absolute values of couplings is limited by the
reconstruction of the total Higgs width. In all other instances the
limiting factor lies somewhere else, so ratios do not give an
advantage there.\bigskip

One of the most interesting question concerning indirect measurements
is how the accessible mass scales compare to typical mass scales of
direct searches. This is the weak point of such analyses at the LHC,
where with the exception of higher-dimensional QCD
operators~\cite{krauss} the direct and indirect reaches for new
physics are comparable.  Given that at the Higgs factory large
momentum flows through any of the Higgs vertices are unlikely, we can
use a non-linearly realized dimension-6
Lagrangian~\cite{nonlinear,legacy3} to link the projected precision,
$0.2\%$ for the $HZZ$ coupling and $1\%$ for many of the other ones, to
a new physics mass scale,
\begin{alignat}{5}
|\Delta_x| &\approx \left| \frac{g^2 v^2}{\Lambda^2} \right| < 
\begin{cases} 
  10^{-2} \qqqquad \quad \; \dfrac{\Lambda}{g} > 10 v = 2.5~\tev 
\\[4mm] 2 \cdot 10^{-3} \qqqquad \dfrac{\Lambda}{g} > 22 v = 5.5~\tev \; .
\end{cases}
 \end{alignat}
For new physics in the electroweak sector this energy range will
hardly be covered by the LHC, with the sole exception of a weak gauge
boson directly produced.  In case of loop effects the typical mass
scale $\Lambda$ will be reduced by a factor $\sqrt{4 \pi}$, still well
above the LHC reach for example in the case of supersymmetric
electroweakinos.\bigskip

While we were in the final stage of our paper, a similar study
indicated that for a linearly realized dimension-6 Lagrangian an
increase in collider energy is more significant~\cite{new}.  We do not
expect this difference to be linked to momentum-dependent couplings in
$W$-fusion, because the corresponding virtuality is known to be
determined by the $W$ and Higgs masses~\cite{virtuality}.  In line
with Ref.~\cite{new}, we speculate that the difference between the two
analyses is explained by inherent strong correlations in the linearly
realized dimension-6 model, particularly between the operators
affecting the $VVH$ couplings. Gauge invariance and custodial symmetry
link the $Z$-couplings and $W$-couplings, where the much higher
statistics for $W$-fusion yields a significant increase in precision.

\section{Summary}
\label{sec:summary}

Measuring the Higgs Lagrangian is the prime motivation for future
$\epem$ colliders. One framework, motivated by a non-linearly realized
dimension-6 Lagrangian, is the measurement of Higgs couplings.  Unlike
at hadron colliders, the $\epem$ environment allows us to measure
Higgs decays into charm quarks and gluons. Even more importantly, we
can directly determine the Higgs width with the help of an inclusive $ZH$ cross
section measurement. This means that $\epem$ data determines all parameters of
the SM-like Higgs Lagrangian independently.  

Two substantially different concepts exist for such a future collider:
the FCCee and CEPC, optimized for delivering a high luminosity, and
the ILC running at a center-of-mass energies up to 
500~GeV. At that energy it probes the $W$-boson fusion production
process combined with all major decay channels as well as
top-quark-associated Higgs production.  A staged ILC running at
250~GeV collider energy and the base-line FCCee are almost identical,
the only difference being the beam polarization. In both cases, the
statistical uncertainties will be substantially smaller than the
current theoretical uncertainties.  Therefore, significant progress on
the theory side is vital to exploit the full potential of future
$\epem$ colliders.

For the circular as well as for the linear designs, the typical
experiment-driven coupling uncertainties are going to be below the
per-cent level. In the effective theory interpretation they probe
energy scales well above what is accessible to the LHC. The ILC design
significantly benefits from the $W$-fusion channel, for example in the
Higgs width extraction. A direct measurement of the top Yukawa
coupling, adding to the expected HL-LHC measurement, increases the
precision of all loop-induced Higgs couplings. Possible upgrades of the
ILC design include a significant increase of the luminosity, also just
above the $ZH$ threshold. The ILC program also foresees 
a long term plan going to a center--of-mass energy of 1~TeV. Similarly, a proposed upgrade of the
circular design targets the energy-related shortcomings with a 350~GeV
run.

Altogether, the competition between ILC and FCCee does not see a clear winner.
With its higher center-of-mass energy, the ILC proposal has a mild
advantage on the Higgs couplings to the $W$ and top quark, which then
propagates to the dimension-5 operators.
Albeit using very different collider features, the two designs are still
similar in terms of their Higgs couplings reach.




\end{document}